\documentclass[amsmath,amssymb, twocolumn]{revtex4-1}

\usepackage{tabularx}
\usepackage{array}
\newcolumntype{L}{>{\centering\arraybackslash}m{2cm}}
\usepackage{graphicx}
\usepackage[usenames,dvipsnames]{color}
\usepackage{latexsym}
\definecolor{orange}{cmyk}{0,0.5,1,0}
\begin{document}
\title{Patterns in Illinois Educational School Data}
\author{Cacey S. Stevens$^1$, Michael Marder$^2$, and Sidney R. Nagel$^1$}
\affiliation{$^1$The James Franck Institute and Department of Physics,
The University of Chicago, Chicago, Illinois 60637, USA}
\affiliation{$^2$Department of Physics, The University of Texas at Austin, 78712 USA}
\date{\today}

\begin{abstract}
We examine Illinois educational data from standardized exams and analyze
primary factors affecting the achievement of public school
students. We focus on the simplest possible models: representation of
data through visualizations and regressions on single variables. Exam scores are shown to depend
on school type, location, and poverty concentration. 
For most schools in Illinois, student test scores decline linearly with poverty concentration.
 However Chicago must be
treated separately.  Selective schools in Chicago, as well as
some traditional and charter schools, deviate from this pattern based on poverty. 
For any poverty level, Chicago schools perform better than those in the rest of
Illinois. Selective programs for
gifted students show high performance at each grade level, most notably at the high school level, when compared to
other Illinois school types. The case of Chicago charter schools is more
complex. In  the last six years, their students' scores overtook
those of students in traditional Chicago high schools. 
\end{abstract}

\maketitle

\section{Introduction}

Each year, public school teachers and administrators face the pressure
of increasing the number of students who meet academic performance
benchmarks.  Standardized exams are a primary measure of this
achievement.  Despite the great differences in support
available to different groups of students, all schools are tested by the same statewide standards.
There is a large amount of
freely available data describing exam results. The Illinois State Board of Education makes aggregate data for all
public schools available each year \cite{ISBE}. Unfortunately the abundance of data is hard to decipher.
We use visualization techniques to display and identify patterns in the data.

Our method of analysis is deliberately simple: scatter plots are employed to detect patterns. 
Our methods are different from multivariate regressions, which are commonly used to study data of this type.
We suspect such multivariate regressions make it difficult for the public to understand the data.
The statistical methods and results require specialized
software and training to
interpret. Thus, before data enter public discourse, they
typically pass through a study or report in which the complex
statistical computations are performed, and their conclusions are deduced for the lay audience.
Such models are not necessary to extract results in this instance. 

Representation by statistical models and analyses is often said to be
  necessary, because raw data provide us with so many numbers that one cannot make sense of them.
However visual images present a
  way to display a large number of data points
  simultaneously, making visualization an important part
  of science, particularly in the study of complex phenomena.
Statistical analysis is often needed to separate true patterns from those due to
  chance. This is particularly true in settings where the number
  of data points is small or the scatter is very large. However, in the cases we will examine, the
  number of data points is so large and the unveiled patterns so
  clear, that elaborate statistical methods are not needed to establish
  them.
The scatter plots have an impact that multivariate regressions lack.
We apply simple regression techniques to show that the patterns found visually are statistically meaningful.

We  explore connections between exam performance in Illinois schools
and various factors, with a particular focus on poverty concentration as  
represented by a conventional proxy: the percent of students who
qualify for lunch at free or reduced price. Poverty concentration of a school is
strongly connected with its performance on standardized exams, as has
been shown before \cite{urban_schools,sirin,bansal,tissington,Marder, bendinelli}. 
In general, as poverty
concentration rises, test scores fall. In most cases, this remains true for charter
schools, which have more autonomy than the traditional public schools
and have been widely promoted as a way for students to escape
unsuccessful local schools.  

We consider additional factors aside from poverty: location and selective programs.
We also study the performance over time. 
Poverty concentration influences performance in Chicago schools differently than the rest of the state.
Additionally, the state has a number of selective schools, where students are chosen by their test scores. 
These schools do not exhibit the patterns relating poverty and test performance. 
The comparisons between selective and open-admissions programs are most striking for high schools. 
Lastly, there are noticeable changes over time, particularly in the performance of charter high schools.

This paper is organized as follows:
Section II describes the types of Chicago public schools and the high-stakes exams in Illinois public education.
In Section III we describe how performance on standardized exams depends on poverty concentration in Chicago compared to the rest of Illinois. The performance of different Chicago school types is also described.
The achievement on measures of college readiness is discussed in Section IV.
In Section V, we show how Chicago schools have been changing over time.
We conclude in Section VI.

\section{Public Education in Illinois}

According to the U. S. Census Bureau, Chicago is by far the largest city in Illinois with about 21\% of the state's 12.8 million residents;
the second largest city in Illinois, Aurora, IL, contains only about 1.5\% of the state's residents \cite{census}.
The Illinois State Board of Education oversees performance of public education throughout the state and sets pre-Kindergarten to 12th grade education policies, while local school districts monitor operation of public schools of the area.
The Chicago Public School System is the third largest school district in the United States and includes about 660 elementary and secondary schools.
Chicago holds 17\% of Illinois public schools and serves about 400,000 students \cite{CPSstat}.
Most Chicago schools are conventional neighborhood programs, but there are also several non-traditional school options.
The school types discussed in this article are defined as follows ~\cite{CPS}:

\vspace{0.2cm}
{\bf{Career Academy}}: Students receive career-focused education and hands-on experience in fields such as health, communications, or manufacturing

{\bf{Charter}}: Independently operated schools open to all Chicago students; students are admitted based on lottery if the number of applicants exceeds the available slots

{\bf{Magnet}}: Places importance on one particular subject area, such as international studies or fine arts; students are typically chosen by lottery after submitting an application

{\bf{Military Academy}}: Students wear uniforms and are taught in a structured environment focused on building team and leadership skills; there are no attendance boundaries for application

{\bf{Neighborhood}}: Traditional programs; students who live within the school's residential boundary do not submit an application to enroll in the school

{\bf{Selective Enrollment}}: Academically advanced programs providing accelerated, college preparatory instruction; enrollment is based on entrance and/or standardized exam scores; they include regional gifted centers housed within neighborhood or magnet schools, where there are two separate programs within the school

\vspace{0.2cm}

Charter schools receive considerable attention, since they are publicly funded schools that operate without the managing control of the school system \cite{charter}.
These schools have been in place in Illinois since 1996 with the purpose of using innovative practices, such as longer school days, autonomy to hire teachers, or subject-focused curricula, to better serve at-risk children \cite{report}.
As of the 2013-2014 academic year, about 75 percent of Illinois charter schools were in the Chicago Public School system, and these programs served about 49,500 students in Chicago \cite{ISBE}.
Many charter programs have several locations, and the city has about 130 charter campuses \cite{CPSstat}.
The growth of charter schools has been promoted in Chicago as an effort to improve the city's education system
\cite{tribune, renaissance}.
Therefore, in spite of recent education budget cuts and neighborhood school closings \cite{school_closings}, new charter schools are steadily opening each year \cite{tribune}.
Accordingly, the performance of charter schools has been the subject of several reports \cite{MinnReport, Booker, pca, Medill}, whose strongly differing conclusions illustrate the degree to which regression methods alone divide specialists and baffle the public.
The goal of this paper is not to argue the effectiveness of any school type.
Instead we describe powerful ways to analyze the school's performance on standardized exams and suggest that simpler formal methods aid in reaching consensus. 

Standardized exams play an important role in providing the data on which public officials base decisions about educational policy. Starting at 3rd grade, Illinois public school students are required to take standardized exams to evaluate their development during the year.
Exams cover subject areas such as mathematics, reading, science, and writing.
The Illinois State Board of Education (ISBE) is responsible for forming the assessment each year, and the student, teacher, school, district, and state are all reviewed based, in part, on the exam results.
In 2014, Illinois adapted the Common Core State Standards, which assess Math and English language skills that students are expected to know after each grade \cite{CCSS}.

Scores are reported in terms of score ranges determined by the ISBE, and students are graded with one of the following performance levels: Exceed Standards, Meet Standards, Below Standards, or Academic Warning \cite{performance}.
A student passes the exam by meeting or exceeding standards of each subject test.
Achievement scores at the school level are expressed as the percent of students in the school who reach each performance level.
Standardized exams, such as the ACT \cite{act}, are used as a measure of college readiness; there is commonly a minimum required score for scholarships or admission into some selective universities.
Up to the 2014 testing season, the ACT was given to Illinois high school juniors as part of the mandatory Prairie State Achievement Exam (PSAE); the ACT will become optional in 2015 \cite{Rado}.
Table \ref{exam} lists information about standardized math exams in Illinois.

\begin{table*}[!] 
\begin{tabularx}{\textwidth}{X | l |X |X}
Exams & Grades & Description & Scores\\
\hline \hline
Illinois Standards Achievement Test (ISAT) & $3^{rd}$ - $8^{th}$   & Measures student math achievement each year & Performance level assigned based on fixed scale \\
\hline
Prairie State Achievement Exam (PSAE) &$11^{th} $&  Assesses student preparation for post-secondary education and ability to apply math skills to workplace & Scores range from 120 to 200; performance level assigned based on fixed scale \\
\hline
ACT & $11^{th} $ &  Exam commonly used for admissions into four-year colleges and universities & Scores range from 1 to 36 \\
\end{tabularx}
\caption{Description of standardized exams from which scores are used for this analysis.}
\label{exam}
\end{table*}

Benchmarks are set for the education system each year, and scores must be reported for school, district, and state accountability purposes.
Schools are then classified according to success at meeting the targets and may be closed if student performance goals are not met.
New educational policies are also implemented to improve students' scores \cite{renaissance}.
In April 2014, Illinois joined many other states receiving a waiver from the No Child Left Behind law \cite{nclb}, protecting school districts from penalty if testing benchmarks are not met.
This is in exchange for a package of additional reform measures.
It is therefore useful to have ways to display how schools have performed and show the effects of such policies.
All school-level performance data is available from ISBE, including school programs, performance scores, and demographics.
Through data visualization of school scores, we find and interpret patterns among school characteristics.

\section{Poverty and School Performance in and out of Chicago}

\begin{figure}[!] 
\begin{center}
\includegraphics[width=2.5in]{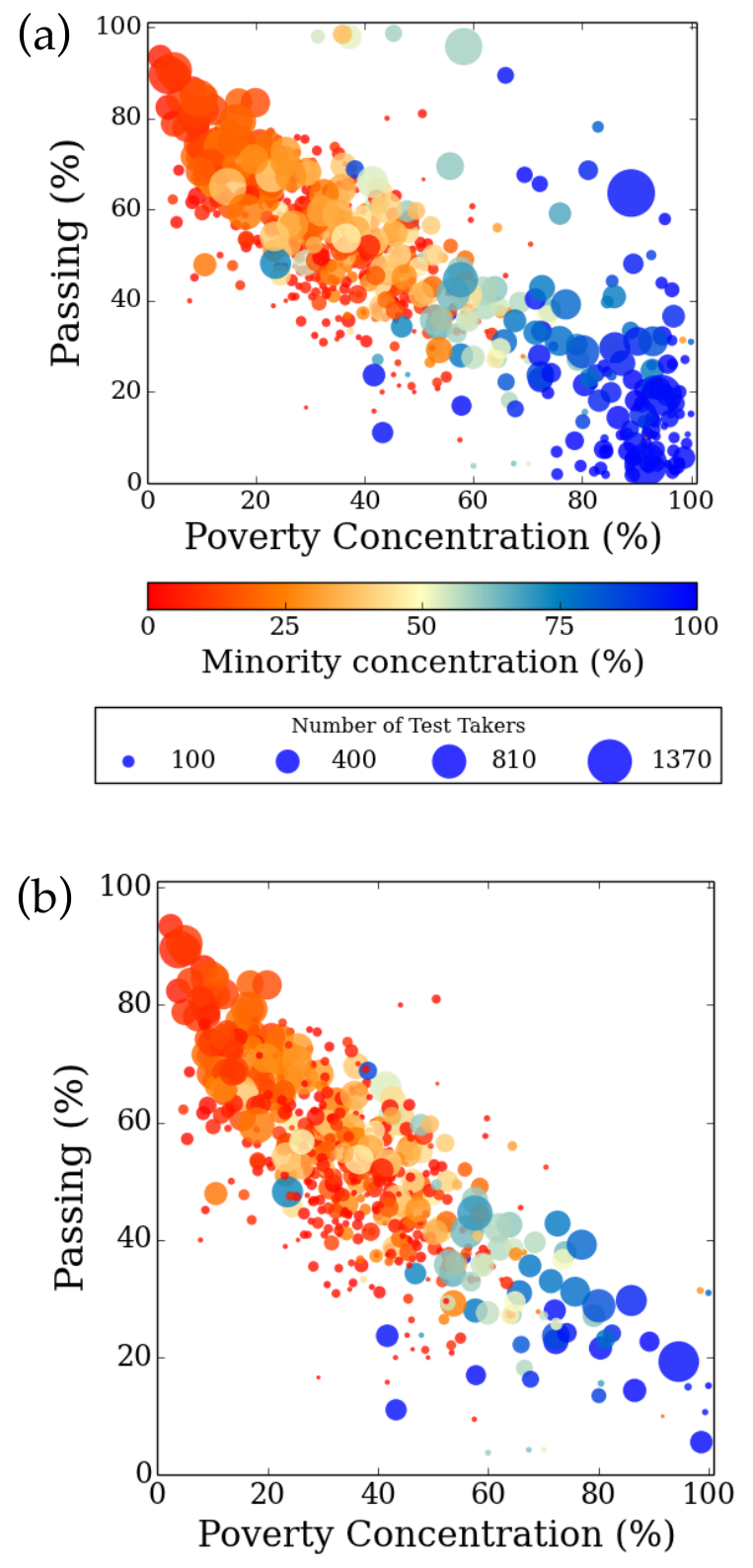}
\caption{(a) Percent of 11th graders at each Illinois school who
  passed the 2013 Math PSAE as a function of poverty concentration. The area of each point represents the
  number of test takers, and the color represents the
  percentage of minority students. (b) Same as (a) but excluding Chicago schools. It becomes particularly clear that Illinois school performance is very tightly
  correlated with poverty concentration and concentration of minority students.}
\label{il_psae}
\end{center}
\end{figure}

Figure \ref{il_psae}(a) displays the performance of 670 Illinois public high
schools as a function of poverty concentration. 
In this case, the measure of school achievement is given as the percentage of students in the school who met or exceeded ISBE standards of the 2013 Mathematics PSAE, an exam given to high school juniors \cite{SPon,psae}.
Poverty concentration is defined as the percent of students who received lunches at no or reduced cost based on the maximum family income requirement.
This is a standard measure of poverty in schools.
For the 2013-2014 academic year, a family of four with an annual household income below \$43,568 qualified for the free/reduced lunch program \cite{lunch}. 
Low-income families choosing not to sign up for the lunch program are not represented.
Two additional factors are expressed in this visualization by the color and size of each circle: the percent of minority students (African American and Hispanic) and the number of test takers at each school, respectively.
Several schools were excluded from the data either for not completely reporting information or for reporting 0\% of students passing. (Waukegan, IL area schools are not displayed due to a discrepancy in 2013 poverty concentration information.)
Data for PSAE Reading and Science exam scores follow similar patterns as described here. 
For brevity, only Math exam performance is shown in this paper.

Performance is strongly connected with concentration of low-income students, as shown by the significant decline in the percent of students passing with increasing poverty concentration. 
Within many schools in low-income neighborhoods (those with greater
than 80\% poverty concentration), less than 50\% of students pass the
exam. 
There is also a connection with minority concentration; the coloring
on the graph shows a strong correlation between poverty concentration and minority concentration.
In 2013, 65\% of White Illinois students passed the exam compared to only 36\% of Hispanic students and 21\% of African American students \cite{ISBE}. 

There are many outliers from the overall decline of performance with poverty shown in Fig. \ref{il_psae}(a).
However, when we
remove all schools located in Chicago, as shown in Fig. 1(b), the remainder of the state's schools present
a remarkably tight relationship between poverty concentration and
passing fraction. For example, there are almost no schools
with more than 80\% poverty concentration and more than 40\% students
passing, as there are no schools with less than 20\%
poverty concentration and less than 40\% students passing.

\begin{figure}[!] 
\begin{center}
\includegraphics[width=2.5in]{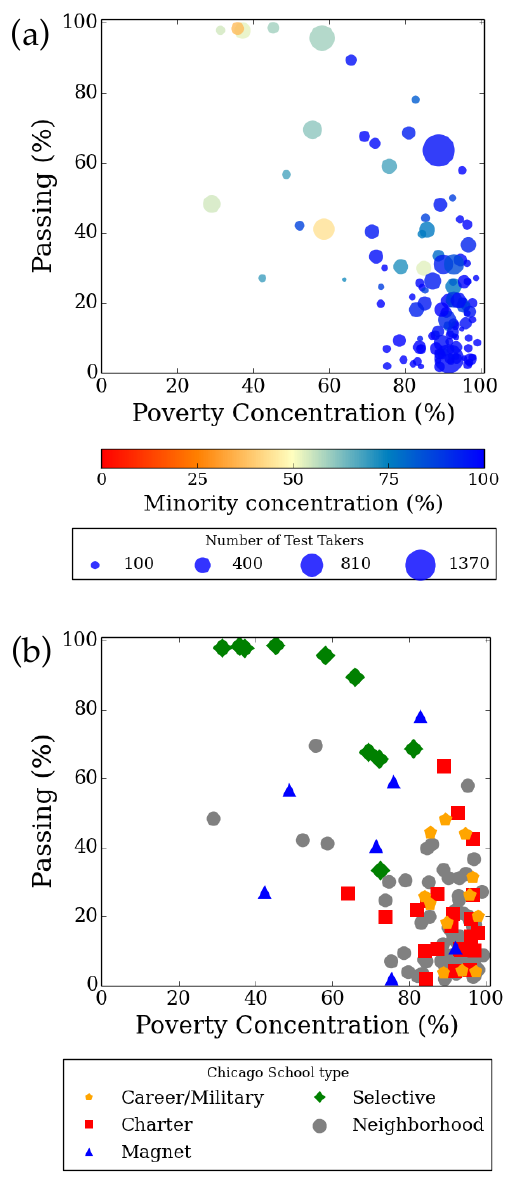}
\caption{(a) Percent of 11th graders at Chicago schools only who
  passed the 2013 Math PSAE as a function of poverty concentration. The size and color of the points represent the number of test takers and percent of minority students respectively. (b) Chicago schools are again presented;  in this case, symbols indicate the type of school as given in the table.}
\label{chicago}
\end{center}
\end{figure}

 When only Chicago schools are considered, we find that the relationship
between poverty concentration and passing fraction is possible to
break.
Figure \ref{chicago}(a) shows the performance of 115 Chicago schools. 
They clearly do not show the same
association with poverty as in Fig. \ref{il_psae}(b). 
Most schools have a poverty concentration greater than 50\%, and among
these schools we see significant variations in performance.
For comparison, Aurora, the second largest city in Illinois, only has four high
schools, and these schools all fit well on the trend shown in Fig \ref{il_psae}(b).
Schools in the areas surrounding Chicago, including Will, Lake, Dupage,
and Cook counties, also fit Fig. \ref{il_psae}(b) well. Only Chicago is different.

\begin{figure*}[!] 
\begin{center}
\includegraphics[width=5.9in,height=4.5in]{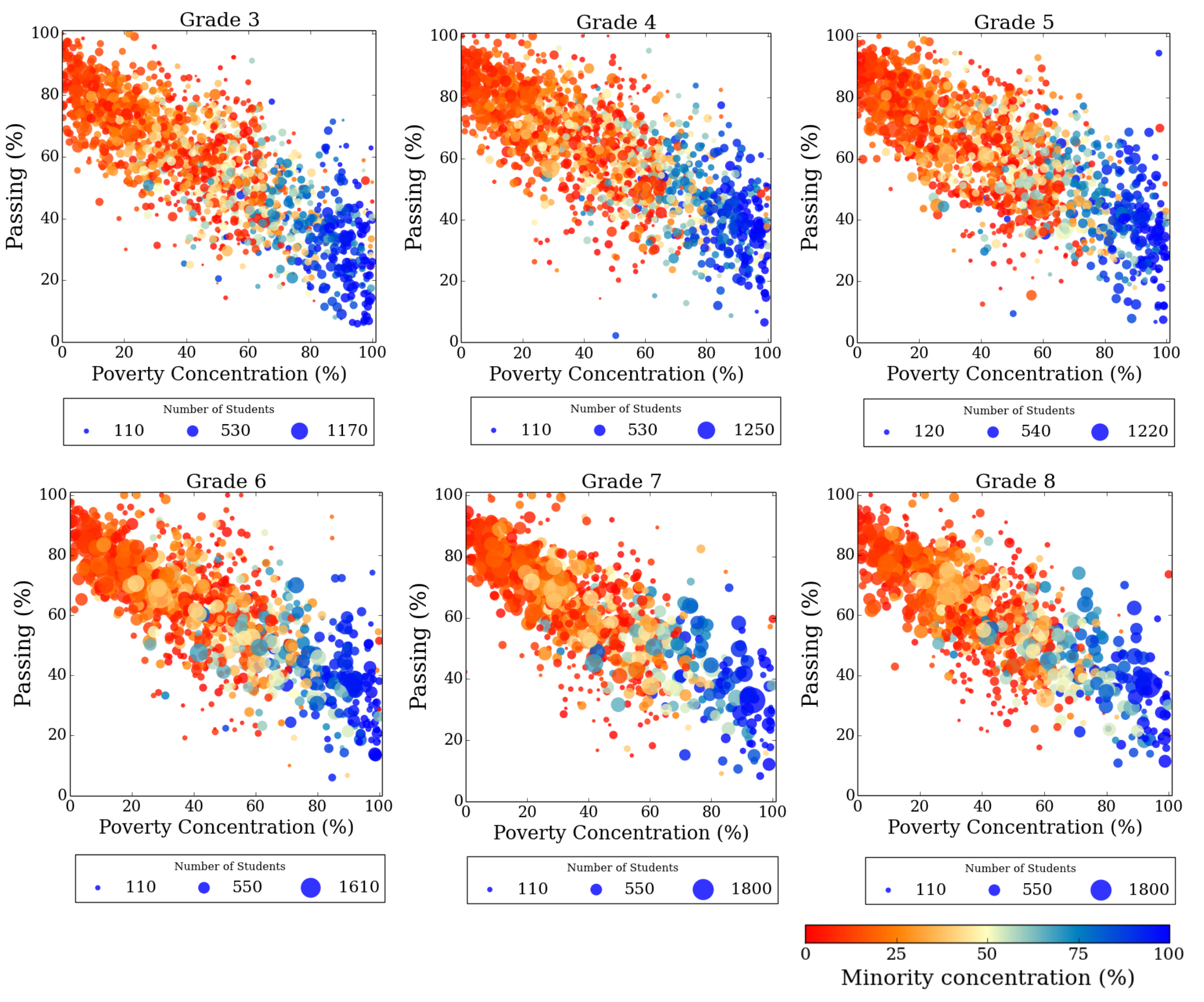}
\caption{Illinois school performance, excluding Chicago schools, on the 2013 Math ISAT at grades 3 through 8. The area of each point represents the
 total number of students tested at each school, and the color represents the
  percentage of minority students in the school. Performance declines with increasing poverty concentration.}
\label{grade}
\end{center}
\end{figure*}

Chicago has several types of programs, as highlighted in Fig. \ref{chicago}(b): career or military, charter, magnet, selective enrollment, and neighborhood.
There is a fairly large spread in achievement among neighborhood, career/military, and
charter schools, though most have less than 50\% passing.
One exception is the Noble Street Charter System where 89\% are considered low-income and 63.6\% of the 1,585 tested students passed the PSAE.
The seven highlighted magnet schools also vary significantly.
Von Steuben High School, with 59\% passing at 76\% poverty concentration, and Devry Advantage High, with 78\% passing at 83\% poverty concentration, perform better than anticipated based on the performance-poverty concentration trend.
Two magnet programs show surprisingly low performance scores.
Administrators of Crane Medical Prep High School,  at 2\% passing,
attribute this to reporting error, while Clark Academic Prep, which
fits the pattern in Fig \ref{il_psae}(b) with 86\% poverty concentration and  10\%
passing, is now on probation. 
As one might expect, the selective enrollment programs, which have competitive admissions and accelerated instruction, have results superior to all other Chicago schools, with several having over 90\% of students passing the Math PSAE.
Regional gifted centers, smaller programs housed within neighborhood or magnet schools, are featured as selective enrollment programs.
For instance, Morgan Park High school has an international studies gifted program within the traditional school and is noted in Fig. \ref{chicago}(b) as a selective enrollment program with 37\% passing.

Proceeding to examine lower grades, we find a very similar set of
patterns: school performance outside of Chicago clusters tightly
around a line where performance declines as poverty rises.
Chicago schools show more complex patterns, which depend, in part, on school type.
In grades 3 through 8,  students' accomplishment is measured
through the ISAT, and, as with the PSAE, ``passing'' is equivalent to
meeting or exceeding exam standards. 
For each grade level, we find the percent of students passing the 2013 exam
at Illinois schools, excluding Chicago, as a function of
poverty concentration; this is shown in Fig. \ref{grade}. 

Since passing fractions of Illinois schools outside of Chicago clearly show a linear relationship with poverty concentration, we can perform simple regressions to obtain the best fit line $f_i$ to describe these patterns.
The quality of the fit is depicted by its
standard deviation $\sigma$, the amount of variation from the line.
The regression line equations and their corresponding $\sigma$ are given in Table \ref{difference}.

\begin{table}
\begin{tabular}{c|c|c }
Grade & Regression Line & $\sigma$\\
\hline \hline
3 &$f= -0.56c_i + 83.7$  & 11.9\\
4 &$f= -0.51c_i+ 86.0$&  12.2\\
5 &$f= -0.52c_i + 85.2$ &  12.4\\
6 &$f= -0.53c_i + 85.5$&   12.5\\
7 & $f= -0.54c_i + 82.9$& 12.1\\
8 &$f= -0.54c_i + 82.5$&  12.1\\
11& $f= -0.67c_i + 76.5$&  10.6\\
\end{tabular}
\caption{Regression lines for passing fraction $p_i$ versus poverty
  concentration $c_i$ plots of each grade level (Figs. 1(b) and 3), and the respective standard deviations $\sigma$. Data are for 2013 Mathematics exams and
  apply to Illinois schools outside of Chicago.}
\label{difference}
\end{table}

\begin{figure}[!] 
\centering
\includegraphics[width=2.44 in]{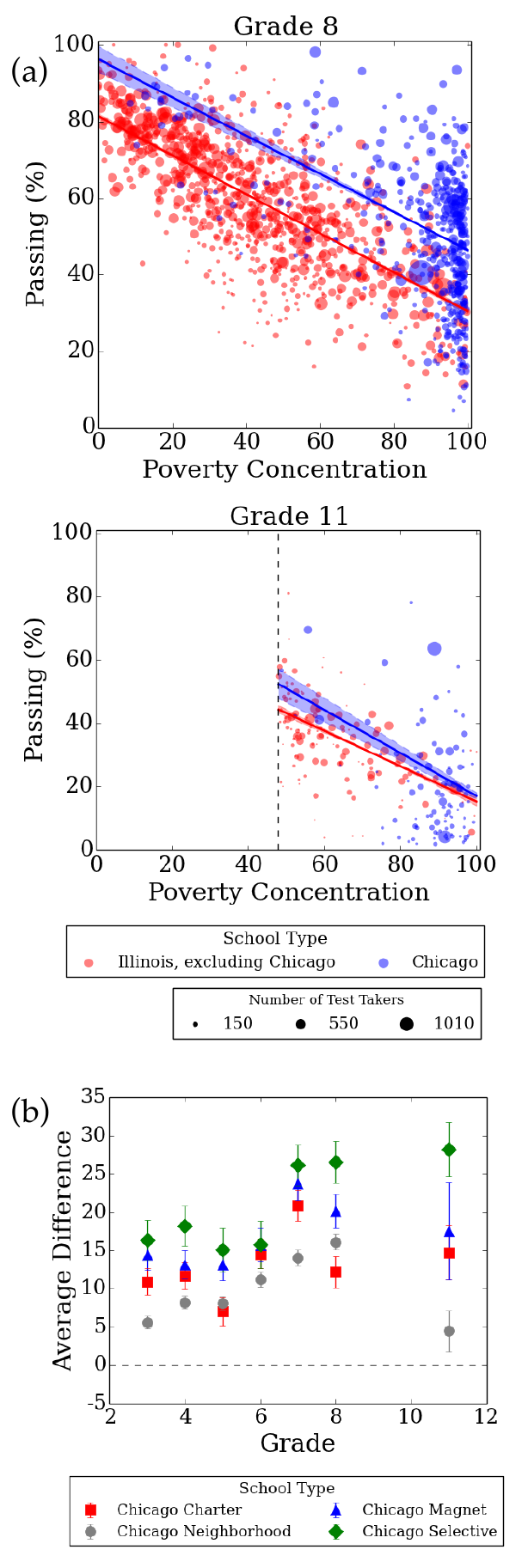}
\caption{(a) Math performance scores of Chicago schools (blue) are compared to the fit to the data of other Illinois schools (red) at $8^{th}$ and $11^{th}$ grades. Transparent bands indicate the standard uncertainty of each line. (b) Average differences of 2013 Math exam performance at Chicago
  school programs when compared to Illinois schools of smaller
  communities. The average difference $\Delta$ of each school category is determined by  $\Sigma N_i (p_i-f_i(c_i))/\Sigma N_i$, where $p_i$ is the percent passing at each school, $f_i$ is the expected performance at that grade level based on the poverty concentration $c_i$ of the school, and 
  $N_i$ is the number of students in each school of that category. Table \ref{schools} gives the number of schools represented.}
\label{deviation}
\end{figure}

We continue to a comparison of performance at schools in and out of
Chicago. 
Figure \ref{deviation}(a) gives examples of how Chicago area schools, excluding selective enrollment programs, perform at $8^{th}$ and $11^{th}$ grades compared to those of the rest of the state.
The solid line indicates the best fit line for each set of data, and the shaded region surrounding the line shows its standard uncertainty.
At grade 8, many of the Chicago schools (shown in blue) earned higher scores than the average of non-Chicago programs (shown in red),
though there is a large spread among Chicago schools. 
Similar patterns appear at $3^{rd}$ to $7^{th}$ grades.
However, for $11^{th}$ grade scores, the difference between Chicago and other Illinois schools is not as significant over the full range of poverty concentration.
The performance of Chicago high schools is best examined by considering programs in which at least 48\% of students are identified as low-income, a criterion that applies to 103 of the 105 career, charter, neighborhood, magnet, and military schools represented.
Since data for less than 48\% poverty concentration is so sparse in Chicago, we discard this region from the comparison.
When restricting attention to the range of 48-100\% poverty concentration, we find that Chicago high schools have significantly higher scores than schools in the rest of Illinois, except for the highest poverty schools where the difference vanishes on average.

Figure \ref{deviation}(b) shows how much each type of school in
Chicago deviates from the fit for the rest of the state (dashed line) at $3^{rd}$ through $8^{th}$ and $11^{th}$ grades.
Career and Military schools are excluded.
For $11^{th}$ grade, only schools with poverty concentration higher than 48\% are considered.
We define 
$\Delta_i$ for a school $i$ to be the difference between  the actual
passing fraction $p_i$ and the fraction $f_i$ we would anticipate based upon
the best fit line of Illinois schools (excluding Chicago) appropriate for that grade at poverty concentration $c_i$. 
The value of  $\Delta_i$, is  positive (negative) if the school
performs better (worse) than expected for other Illinois programs. The average difference is weighted by the number of test takers at each school.
For example, the average difference of
Chicago magnet schools is given by 

\begin{equation}
\frac{\sum_{i\in\text{magnet}}(\Delta_i
N_i)}{\sum_{i \in \text{magnet}}N_i} 
\end{equation}
where $N_i$ is the number of
students in each school and the total number of schools is given in Table \ref{schools}. 

\begin{table}[h]
\begin{tabular}{c | c | c | c | c | c}
Grade & Charter & Magnet & Selective & Neighborhood & Illinois \\
\hline \hline
3 & 21 & 35 & 14 & 427 & 1685\\
4 & 21 & 36 & 14 & 422 & 1654\\
5 & 24 & 36 & 14 & 421 & 1571\\
6 & 23 & 37 & 14 & 405 & 1107\\
7 & 24 & 38 & 17 & 390 & 913\\
8 & 22 & 37 & 18 & 389 & 911\\
*11 & 22 & 6 & 6 & 63 & 129\\
\end{tabular}
\caption{Number of schools represented by each point of Fig. 4(b). * At grade 11, only schools with more than 48\% concentration of low-income students are considered. }
\label{schools}
\end{table}

At every grade level, all Chicago school types, including neighborhood schools,
perform better, on average, than expected based on achievement of the rest of the state.
Neighborhood schools have increasingly better performance up to grade 8.
 At $11^{th}$ grade, Chicago's neighborhood programs score slightly above the rest of the state though less pronounced than at lower grades. 
Additionally, charter school systems perform below other Chicago schools at grades 5 and 8, and magnet programs lie consistently above charter and neighborhood schools.
Selective enrollment schools, not
surprisingly, have the highest performance at all grades.
This is especially striking at $11^{th}$ grade where selective enrollment programs show a significant jump, while neighborhood schools decline.
In high school, most gifted programs are full-site centers, meaning
that all students follow the more rigorous curriculum.
In contrast, ten of the
elementary level selective enrollment programs are regional gifted
centers. High school students in selective programs must also have demonstrated high scores on previous standardized exams (ISAT).
On the other hand, elementary students are admitted based on a school's entrance exam and not necessarily performance on standardized state exams.
This likely contributes to the increased difference at $11^{th}$ grade compared to lower grades in gifted programs.

\section{Measure of College Readiness}

The ACT measures college readiness at high schools.
Accordingly, we note whether, on average, students of Illinois schools are attaining satisfactory ACT scores.
Up to the 2014-2015 academic year, Illinois $11^{th}$ graders took the ACT for free as part of the mandatory state administered exams offered to public school students \cite{Rado}.
Therefore, the ACT is the key exam for college admissions, as opposed to the SAT.

\begin{figure}[h] 
\centering
\includegraphics[width=2.6 in]{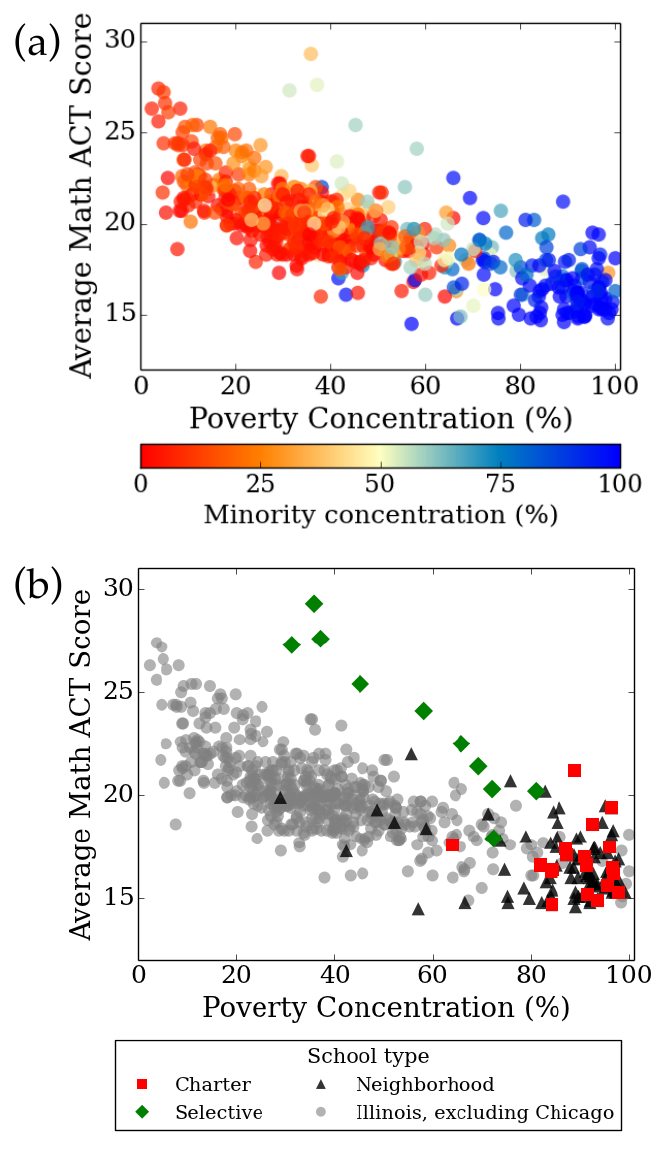}
\caption{Average Math ACT score of each Illinois school. (a) Math ACT score as a
  function of poverty concentration. The color of the points indicates the percent of minority students at each school. (b) School types are highlighted:
  Chicago charter schools, selective enrollment schools in Chicago,
  neighborhood schools in Chicago, and the rest of Illinois.}
\label{act}
\end{figure}

Figure \ref{act}(a) shows the average Math ACT score of each Illinois high school as a function of poverty concentration.
As with other state standardized exams, ACT Math performance declines with increasing concentration of low-income students.
This is in agreement with a study of Dixon-Roman \textit{et. al.} indicating a correlation between parental income and student's performance on college readiness exams \cite{Dixon-Roman}.
Most schools with less than 20\% poverty concentration have an average score above 20, while, among schools with more than 80\% poverty concentration, few have an average score above 20.
The influence of poverty on college admissions exam scores is not unique to Illinois schools.
For other states that have been examined, poverty is also connected to school performance in Mathematics \cite{Marder}.
Similarly, performance declines with increased concentration of minority students.
 
We study how Chicago schools possibly diverge from this trend, as seen with scores of the PSAE and ISAT.
In Fig. \ref{act}(b), we highlight schools located in Chicago. The symbols represent the different types of programs: Chicago charter, Chicago selective
enrollment, Chicago neighborhood, or Illinois, excluding Chicago. 
All Chicago neighborhood schools cluster at average scores of about 15 to 20, and
there is no significant trend with poverty among neighborhood schools.
The best performing schools are all selective enrollment, college
preparatory schools in Chicago, although performance in these schools does decline with poverty.
With an average score of 17.9, Morgan Park High, a regional gifted center, is the lowest among gifted programs.

Previous analyses show that attending a Chicago charter high school
may have positive effects on ACT scores \cite{Booker}. 
It is often argued that charter schools improve student achievement due to innovative methods of instruction and discipline to prepare students for college.
However, our present study 
indicates that scores of most charter schools cluster around similar scores to neighborhood
schools at the same poverty level, as was also noted in a recent report
\cite{Medill}. There is an exception: the Noble
Street Charter System with an average score of 21.2. 
Note that ISBE reports only one score of all campuses of charter school systems, such as Noble Street Charters. 
According to a separate performance report from the Chicago Public School System, the average scores on the 2013 Math ACT among Noble Street schools ranged from 19.2 at the Rowe-Clark campus to 23.7 at the UIC campus \cite{CPSdata}.

\section{School Performance Over Time}

\begin{figure}[!] 
\includegraphics[width=2.6 in]{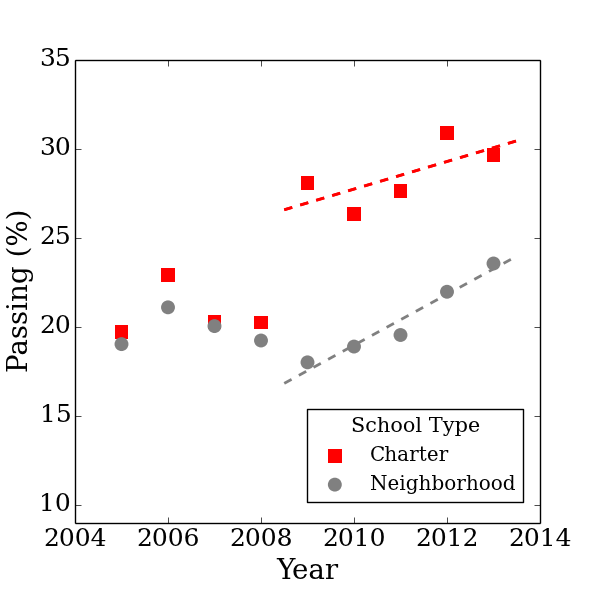}
\tiny
\begin{tabular}{|c | c | c | c | c | c | c | c | c | c |}
\hline
  & 2005 & 2006 & 2007 & 2008 & 2009 & 2010 & 2011 & 2012 & 2013 \\
\hline \hline
 & \multicolumn{9}{c}{Number of School Systems}\\
\hline
Charter & 7 & 8 & 9 & 11 & 13 & 14 & 15 & 18 & 22\\
Neigh. & 53 & 58 & 58 & 67 & 62 & 65 & 67 & 64 & 64\\
\hline
 & \multicolumn{9}{c}{Number of Students Tested}\\
\hline
Charter & 735 & 882 & 1026 & 1131 & 1869 & 2489 & 3926 & 4688 & 5305\\
Neigh. & 11287 & 11140 & 11807 & 14099 & 13640 & 12832 & 14452 & 13987 & 13297\\
\hline
\end{tabular}
\caption{Percent of Chicago charter and traditional high school students who
  passed the Math PSAE each year. There was a performance shift in
  2009 after which a greater percent of charter students passed. Neighborhood schools have improved at about 1.5\% per year from 2009-2013, compared to 0.8\%/year among charter schools, as determined from the slopes of each fit. The
  table gives the number of schools and the number of students
  tested each year in those schools.} 
\label{psae_long}
\end{figure}

Thus far we have only considered school performance in 2013, comparing various types of educational programs at each grade level.
In this section we address the evolution of these results over time. 
For example, the number of charter schools in Chicago has grown consistently as new schools and locations open.
Accordingly, we ask if there has been an overall improvement of the charter school system as it grew.
Figure \ref{psae_long} shows the total percent of $11^{th}$ grade students in
Chicago charter and neighborhood high schools (excluding selective enrollment, magnet, career, and military schools) who passed
the Math PSAE each year. 
The numbers of schools and students represented are given in the table of Fig. \ref{psae_long}.
The percent of students passing is obtained by the fraction of the sum of
all students in each charter or high school who met or exceeded exam
standards to the total number of students tested in those schools
each year:

\begin{equation}
\frac{\Sigma(p_i*N_i)}{\Sigma N_i} 
\end{equation}

\noindent
where $p_i$ is the percent passing and $N_i$ is the number of students tested at each school of the program type (charter or traditional) in each year. The concentration of low-income students is not considered in this comparison.
    
The percentage of students passing remained at only about 18\% to 33\% at Chicago neighborhood and charter programs from 2005 to 2013, compared to about 60\% of Illinois school students of other communities (due to the lower concentration of low-income students).
Up to 2008, charter schools consistently performed at the same level as other high
schools in the percent of students passing the exam.
However, in the last several years, there has been a shift in performance; between 25\% and 30\% of charter students passed the exam each year since 2009.
Charter school scores have increased by about 0.8\% per year from 2009 to 2013.
Although neighborhood high schools in Chicago have improved by about 1.5\% each year, the traditional programs have
not reached the performance of charter schools. 

Some charter programs have been criticized for their high attrition rates \cite{educweek}; charters expel poorly performing students at a higher rate than at other public schools, resulting in a boost in test scores.
Chicago Public Schools recently released data that support the claim \cite{tribune_expulsions}.
 The expulsion rates of 2012-2013 at Chicago charters was 0.61\% compared to only 0.05\% at traditional schools.
 Some charter programs noted for high academic performance have expelled 1.5 - 4.8\% of the student body during the year \cite{CPSdata}.
Here we provide a rough estimate of the impact of high expulsion rates on Math PSAE performance:
 We assume the extreme case where all expelled students would not have passed the exam and were thereby transferred to a neighborhood school.
 If those students were not expelled and were instead tested in a charter program, the resulting percent passing in 2013 among charter students would have been 29\% and would have increased among neighborhood students to 24\%.
Thus, the higher scores of charter programs are not fully explained by their higher expulsion rates.

The number of charter systems has increased in the last nine years from 7 in 2005 to 22 in 2013, not including new campuses of established systems; there were 54 charter high school campuses as of the 2012 academic year. 
Note that charter systems with several locations may report scores and demographics as one school.
In 2008-2009, the number of charter school $11^{th}$ graders grew by 65\%, while some traditional public schools were closed or relocated as part
of the Renaissance plan of Chicago Public Schools
\cite{renaissance}. This must have contributed to the shift apparent in 2009.
Even if charter schools pulled stronger students away from neighborhood schools after 2009, the data in Fig. 4(b) indicate the net result for Chicago has been positive.
When considering the scores of each charter school separately, we find
that most schools do not have a major change in the percent of
students passing the exam over several years. Schools that start with strong
scores typically remain at the same level, and schools with lower
scores do not usually improve. Over the last five years, there have been new charter programs
opening, and some of the schools have better performance scores within
the first year than established traditional high schools.  Accordingly the overall achievement of charter schools on the $11^{th}$ grade Math exam has been above that of neighborhood schools.

\section{Conclusions}
We have analyzed standardized exam performance at Illinois schools from
2005 to 2013, making use of public data that anyone can
access. 
Through data visualization methods commonly used in science, we reveal clear patterns in Illinois education.
We have shown that performance at all schools outside of Chicago, depends
very consistently upon poverty concentration. 
Outside Chicago, no schools in Illinois with a high concentration of low-income students have performance scores as great as
schools in better-off communities with a poverty-concentration less than 20\%. 
In Chicago, the story is more complicated. 
Schools in the city have higher scores, on average, than the rest of the state at each grade and poverty level.
Performance of Chicago schools depends, in part, on the type of program, such as programs for gifted students and networks of charter schools.
Selective enrollment programs consistently have the highest exam achievement compared to other public schools.
This is especially true at $11^{th}$ grade.
These programs also excel in measures of college readiness, such as the ACT, though scores are shown to decline with poverty concentration in this case.
In 2013, neighborhood schools performed better than charter schools in grades 5 and 8, and neighborhood and charter high schools have earned similar ranges of performance scores on the ACT.
While schools in the rest of Illinois
displayed a largely static and discouraging dependence
of test performance on poverty, Chicago has improved over the nine year period we examined.
In 2009, charter high schools had a shift in scores from about 20\% to 30\% passing math exams, while traditional schools have gradually improved since 2009.
The overall picture is one in which change in Chicago schooling has produced improvement in test scores for students at all levels of poverty and in all types of schools. 
Meanwhile, in the rest of the state, the dependence of scores on poverty is stronger and is not changing.

\begin{acknowledgments}
We are grateful to Thomas Caswell, Leo Kadanoff, and Andrzej Latka for helpful discussions. C.S. acknowledges support from the NSF Graduate Research Fellowship Program.
\end{acknowledgments}

\bibliographystyle{apsrev}
\bibliography{edILbib.bib}

\end{document}